\documentstyle[12pt,aaspp4,epsf]{article}

\newcommand{\etal}{{\it et al.}}

\newcommand{\kms}{km\,s$^{-1}$}

\newcommand{\kmsyr}{km\,s$^{-1}$\,yr$^{-1}$}
\newcommand{\kmsmas}{km\,s$^{-1}$\,mas$^{-1}$}
\newcommand{\muasyr}{$\mu$as\,yr$^{-1}$}

\newcommand{\msun}{\mbox{M$_{\odot}$}}

\newcommand{\muas}{\mbox{$\mu$as}}
\newcommand{\bq}{\begin{equation}}
\newcommand{\eq}{\end{equation}}
\newcommand{\ho}{H$_{2}$O}

\newcommand{\avgrs}{\mbox{$\langle r_{s}\rangle$}}
\newcommand{\avgax}{\mbox{$\langle \dot{v}_{los} \rangle$}}
\newcommand{\avgpm}{\mbox{$\langle\dot{\theta}_{x}\rangle$}}

\newcommand{\MM}{\mbox{${\cal M}_{7.2}$}}
\newcommand{\dmpc}{\mbox{$D_{6}$}}
\newcommand{\omegas}{\mbox{$\Omega_{s}$}}
\newcommand{\is}{\mbox{$i_{s}$}}
\newcommand{\pas}{\mbox{$\alpha_{s}$}}

\newcommand{\cit}[1]{$^{\ref{#1}}$}
\newcommand{\citt}[2]{$^{\ref{#1},\ref{#2}}$}
\newcommand{\cittt}[3]{$^{\ref{#1},\ref{#2},\ref{#3}}$}
\newcommand{\citttt}[4]{$^{\ref{#1},\ref{#2},\ref{#3},\ref{#4}}$}

\begin{document}
\vskip 0.6 in
\noindent
 
\title{A 4\% geometric distance to the galaxy NGC4258 from orbital 
motions in a nuclear gas disk}
 
\author{
J.~R. Herrnstein\footnotemark[1]\footnotemark[2],
J.~M. Moran\footnotemark[2],
L.~J. Greenhill\footnotemark[2], 
P.~J. Diamond\footnotemark[1]\footnotemark[3],
M. Inoue\footnotemark[4],
N. Nakai\footnotemark[4], 
M. Miyoshi\footnotemark[5], 
C. Henkel\footnotemark[6],
A. Riess\footnotemark[7]
}
\footnotetext[1]{National Radio Astronomy Observatory, PO Box O, Socorro, NM 
87801}
\footnotetext[2]{Harvard-Smithsonian Center for Astrophysics, Mail Stop 42, 
60 Garden Street, Cambridge, MA 02138}
\footnotetext[3]{Merlin and VLBI National Facility, Jodrell Bank, 
Macclesfield, Cheshire SK11 9DL, U.K.}
\footnotetext[4]{Nobeyama Radio Observatory, National Astronomical Observatory, 
Minamimaki, Minamisaku, Nagano 384-13, Japan}
\footnotetext[5]{VERA Project Office, National Astronomical Observatory, Mitaka, 
Tokyo, 181-8588, Japan}
\footnotetext[6]{MPfIR, Auf dem Hugel 69, D-53121, Bonn, Germany}
\footnotetext[7]{Department of Astronomy, University of California at Berkeley, 
Berkeley, CA 94720}

\slugcomment{****** To Appear In Nature; Draft of May 18; 1300 words + 3 figs + 2 eqs ******}
 
{\bf The accurate measurement of extragalactic distances is a central challenge of 
modern astronomy, being required for any realistic description of the age, geometry 
and fate of the Universe. The measurement of relative extragalactic distances has 
become fairly routine, but estimates of absolute distances are rare\cit{jacoby92}. 
In the vicinity of the Sun, direct geometric techniques for obtaining absolute 
distances, such as orbital parallax, are feasible, but heretofore such techniques 
have been difficult to apply to other galaxies. As a result, uncertainties in the 
expansion rate and age of the Universe are dominated by uncertainties in the absolute 
calibration of the extragalactic distance ladder\cit{madore99}. Here we report a 
geometric distance to the galaxy NGC4258, which we infer from the direct measurement 
of orbital motions in a disk of gas surrounding the nucleus of this galaxy. The distance 
so determined - $7.2\pm0.3$\,Mpc - is the most precise absolute extragalactic distance 
yet measured, and is likely to play an important role in future distance-scale 
calibrations.}

NGC4258 is one of 22 nearby AGN known to possess nuclear water masers (the microwave 
equivalent of lasers).  The enormous surface brightnesses ($\ga10^{12}$\,K), small 
sizes ($\la10^{14}$\,cm), and narrow linewidths (a few\,\kms) of these masers make 
them ideal probes of the structure and dynamics of the molecular gas in which they 
reside.  VLBI observations of the NGC4258 maser have provided the first direct 
images of an AGN accretion disk, revealing a thin, subparsec-scale, differentially
rotating warped disk in the nucleus of this relatively weak Seyfert 2 
AGN~\citttt{watson94}{miyoshi95}{greenhill95a}{herrnstein96}.  Two distinct 
populations of masers exist in NGC4258.
The {\it high-velocity} masers amplify their own spontaneous emission and are offset 
$\pm1000$\,\kms\ and 4.7-5.1\,mas (0.16-0.28\,pc for a distance of 7.2\,Mpc) on either
side of the disk center.  The beautiful Keplerian rotation curve traced by these masers 
requires a central binding mass ($M$), presumably in the form of a supermassive 
black hole, of $(3.9\pm0.1)\times10^{7}(D/7.2\mbox{ Mpc})(\sin\is/\sin82)^{-2}$ solar 
masses (\msun) where $D$ is the distance 
to NGC4258 and \is\ is the disk inclination. Because the high-velocity masers lie in
the plane of the sky, they should to first order remain 
stationary as the disk rotates. The {\it systemic masers}, on the other hand, are 
positioned along the near edge of the disk and amplify the background jet emission 
evident in Figure~1~\cit{herrnstein97b}.  A fundamental and as-yet untested 
prediction of the maser disk model is that the systemic masers should drift with 
respect to a fixed point on the sky by a few 10\,\muasyr\ as the disk rotates at 
$\sim1000$\,\kms.

\begin{figure}[htbp]          
\plotone{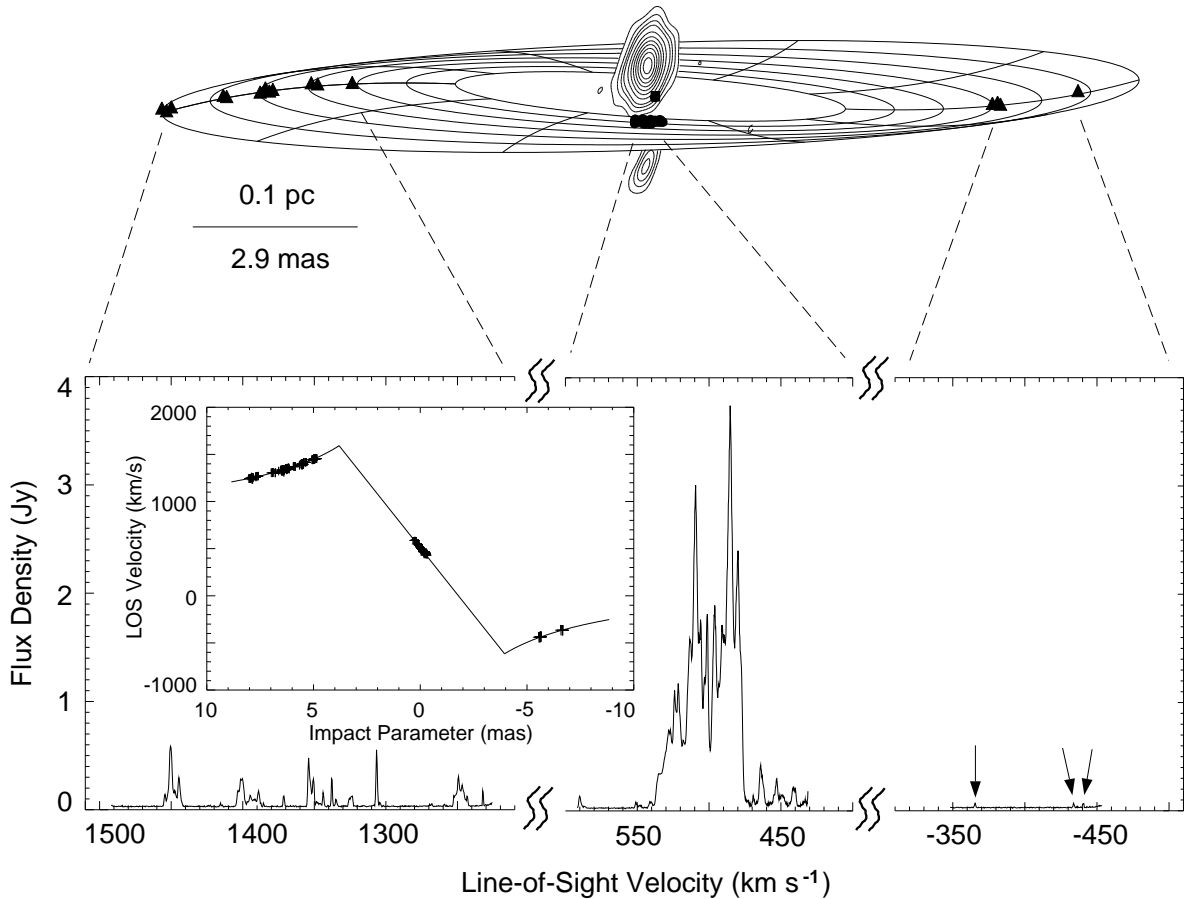}
\caption {See captions at end of text.}
\label{fg:fig1}
\end{figure}

\begin{figure}[htbp]          
\plotone{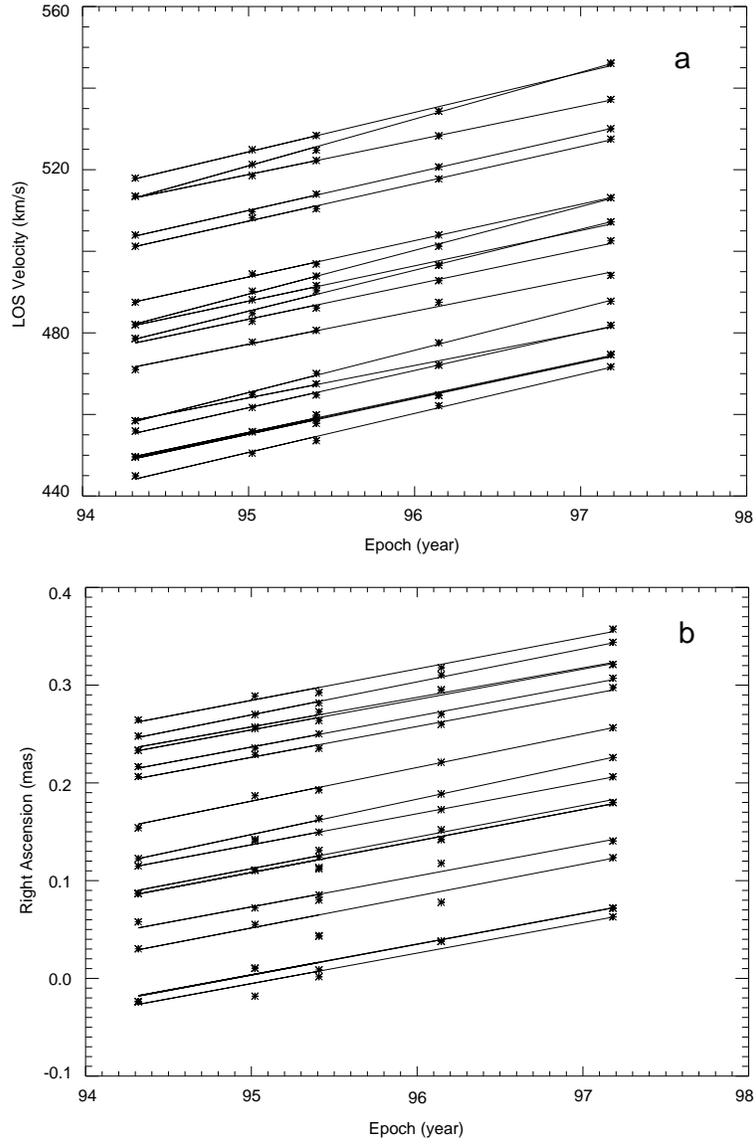}
\caption {See captions at end of text.}
\label{fg:fig2}
\end{figure}

NGC4258 was observed at 4--8 month intervals between 1994 and 1997 with the VLBA of 
the NRAO in order to search for the expected motions.  Since the maser emission is 
essentially continuous across the envelope of systemic maser emission, we are forced 
to rely on structure in the systemic spectrum to isolate and track individual maser 
features.  The assumption that distinct peaks in the spectrum correspond to individual 
clumps of gas is justified by the successful tracking of maser accelerations in 
single-dish monitoring programs, and leads to an identification of 20--35 potentially 
trackable systemic masers in each epoch.  These features are too densely packed in 
position and velocity to be individually tracked with any reliability.  However,
the resolution and sampling are sufficient to detect bulk rotation in the system, 
and we have developed a Bayesian pattern-matching analysis tool to track inter-epoch 
shifts in the positions and velocities of the systemic masers, as a whole\cit{myphd}. 
The analysis assumes the systemic masers are randomly and narrowly scattered about an 
average radius, \avgrs, of 3.9\,mas, as indicated by the global disk-fitting analysis
(see Figure~1). The precise magnitude of the radial scatter is set so as to maximize 
the overall likelihood of the tracking analysis.  In order to evaluate the likelihood 
of a given bulk proper motion (\avgpm) or acceleration (\avgax), the pattern-matching 
procedure must compute the likelihood that each individual maser has in fact moved by 
\avgpm\ or \avgax.  This leads to robust estimates for the ``trackability'' of each 
maser.  Figure~2 shows the best-fitting acceleration and proper motion tracks for 
the most reliably trackable systemic masers.  Figure~3 shows the overall probability 
density functions 
(PDFs) for \avgax\ and \avgpm.  The PDFs indicate a highly significant detection of 
bulk motion in the disk and from them we conclude $\avgax=9.3\pm0.3$\,\kmsyr\ and 
$\avgpm=31.5\pm1$\,\muasyr, where these and all subsequent uncertainties are $1\sigma$
values.  The latter result is consistent with expectations and is the first detection 
of transverse motion in the NGC4258 accretion disk.  We note that the pattern-matching 
algorithm has been verified on a number of simulated datasets with feature densities 
and spectral and spatial resolutions comparable to those of the true data.

In order to convert the maser proper motions and accelerations into a geometric distance, 
we express \avgpm\ and \avgax\ in terms of the distance and four disk parameters: 
\bq
\avgpm=31.5\left[\frac{\dmpc}{7.2}\right]^{-1}\left[\frac{\omegas}{282}\right]^{1/3}\left[\frac{\MM}{3.9}\right]^{1/3}\left[\frac{\sin\is}{\sin 82.3^{\circ}}\right]^{-1}\left[\frac{\cos\pas}{\cos 80^{\circ}}\right]\mbox{~~~\muasyr},
\label{eq1}
\eq
and
\bq
\avgax=9.2\left[\frac{\dmpc}{7.2}\right]^{-1}\left[\frac{\omegas}{282}\right]^{4/3}\left[\frac{\MM}{3.9}\right]^{1/3}\left[\frac{\sin\is}{\sin 82.3^{\circ}}\right]^{-1}\mbox{~~~\kmsyr}.
\label{eq2}
\eq
Here \dmpc\ is the distance in Mpc, \pas\ is the disk position angle (East of North) 
at \avgrs, and \MM\ is $M/D\sin^{2}\is$ as derived from the high-velocity rotation curve 
and evaluated at $D=7.2$\,Mpc and $\is=82.3^{\circ}$ (in units of $10^{7}$\,\msun). 
$\omegas\equiv(G\MM/\avgrs^{3})^{1/2}$ is the projected disk angular velocity at \avgrs\ 
as determined by the slope of the systemic position-velocity gradient (in units of 
\kmsmas; see Figure~1). {\it a priori} estimates for each of these disk parameters, 
derived directly from 
the positions and velocities of the masers, are included in the denominators of
each of the terms of equations~\ref{eq1} and \ref{eq2}.

\begin{figure}[htbp]          
\plotone{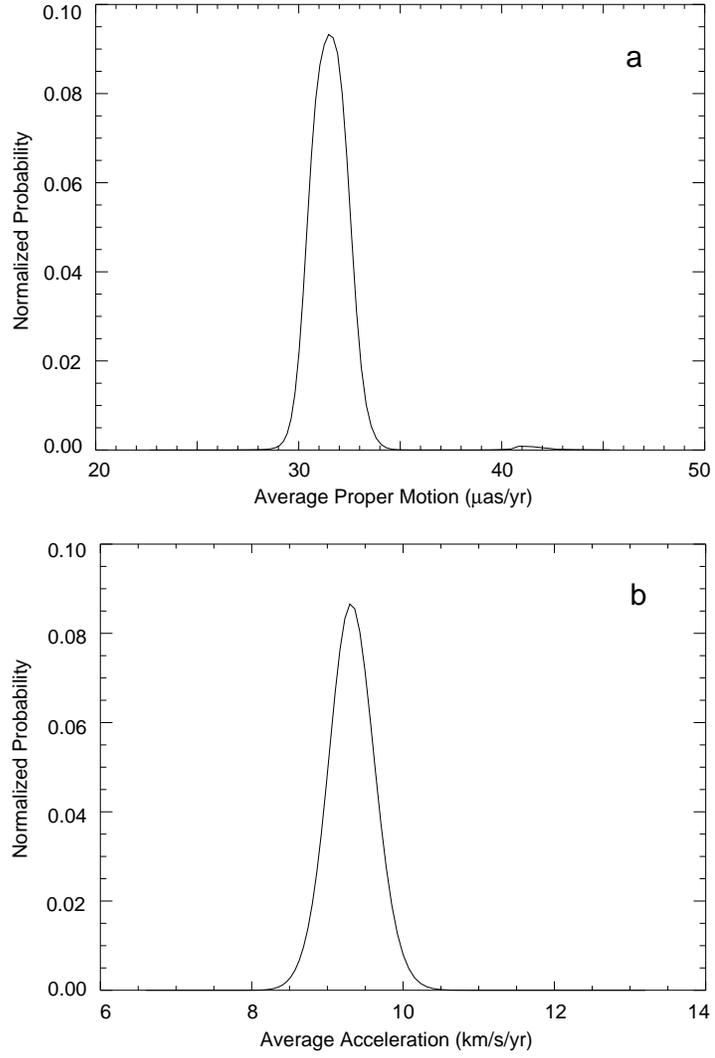}
\caption {See captions at end of text.}
\label{fg:fig3}
\end{figure}

When the {\it a priori} disk parameter estimates are used, the proper motions and 
accelerations yield independent distance estimates, through equations~\ref{eq1} and 
\ref{eq2}, of $7.2\pm0.2$\,Mpc and $7.1\pm0.2$\,Mpc, respectively.  The quoted 
uncertainties are effectively the uncertainties in \avgpm\ and \avgax\ recast in 
terms of distance, and as such are purely statistical in nature.  The excellent 
agreement between the proper motion and acceleration distances for {\it a priori}
values of the disk parameters is an impressive confirmation of the {\it a priori} 
disk model itself, and establishes the NGC4258 Keplerian disk as a fully 
self-consistent, dynamical model incorporating the positions, LOS velocities, 
proper motions, and accelerations of all the NGC4258 masers.  We note that a 
preliminary geometric distance estimate, based on accelerations alone and a single 
VLBA epoch, yielded $D=6.4\pm0.9$\,Mpc~\cit{miyoshi95}.  Hence, the old and 
new distances are consistent at the $1\sigma$ level.  The discrepancy between 
the estimates is ultimately explained by the fact that the original distance 
estimate assumed an average systemic maser radius about 10\% larger than the 
value indicated by the newer data and the more sophisticated disk models.

Uncertainties in the disk parameters contribute to systematic uncertainties 
in the distance estimate. We derive a composite geometric distance estimate using
equations~\ref{eq1} and \ref{eq2}, and from the \avgax\ and \avgpm\ PDFs of Figure~3 
and the {\it a priori} estimates for the disk parameters and their associated 
uncertainties.  {\it The result is a geometric distance estimate of $7.2\pm0.3$\,Mpc,
where the quoted uncertainty now incorporates all statistical 
terms associated with tracking motions in the disk as well as systematics arising
from disk parameter uncertainties.}  A $\sim5$\% uncertainty in \omegas\ is the dominant 
contributor to the latter, producing fractional uncertainties in the acceleration and 
proper motion distances of 6.7\% and 1.7\%, respectively. In total, disk-model systematics 
contribute an additional 0.26\,Mpc (in quadrature) to the distance error budget derived 
from purely statistical considerations.  

The NGC4528 geometric distance is the most precise, 
absolute extragalactic distance measured to date and, being independent of all other 
distance indicators, it represents an important new calibration point for the 
extragalactic distance ladder.  The geometric distance is consistent with pre-existing
H-band Tully-Fisher ($7.1\pm1.1$\,Mpc~\cit{aaronson82}), blue Tully-Fisher 
($7.9\pm1.8$\,Mpc~\cit{richter84}), and luminosity class ($8.4\pm2.2$\,Mpc~\cit{rowan85})
distance estimates, all of which rely on the Cepheid Period-Luminosity relationship 
for absolute calibration.  Most importantly, efforts are underway to determine 
directly a Cepheid distance to NGC4258 from Hubble Space Telescope (HST) observations 
of the galaxy. The NGC4528 geometric distance is presently the most precise means 
for directly calibrating distances obtained by HST Cepheid observations.

We emphasize that the above error budget considers statistical and 
systematic uncertainties within the framework of a thin Keplerian disk in 
which the masers trace the orbital motions of discrete clumps of gas.  As 
always, it is difficult to estimate any additional systematic uncertainties 
that might exist as a result of imperfections in the model itself. The 
potential impact of eccentricity on the distance error budget depends on 
the assumed distribution of accretion disk eccentricities in AGN in general.  
Viscous dissipation within such disks is expected to circularize orbits on 
relatively short timescales, and detailed modeling of the optical emission 
lines from a large sample of AGN suggests eccentricities less than about 
0.5~\cit{eracleous95}.  The eccentricity of the NGC4258 disk is further
constrained by the symmetry of the maser emission about the disk center 
in both position and velocity\cit{herrnstein97b}.  These additional constraints 
lead to an 
expected eccentricity of zero with a probable error of 0.1, a negligible
bias in the distance estimate, and a systematic uncertainty in the distance 
of 0.4 Mpc.  Hence our distance estimate, including this uncertainty in the 
eccentricity, is $7.2\pm0.5$\,Mpc.
Finally, we cannot unambiguously 
rule out contamination of the maser dynamics by some non-kinematical contribution, 
such as traveling density waves within the disk.  However, given 
the complexity (30 systemic masers across $8^{\circ}$ of disk azimuth) and the stability 
($\ga70$\% of the features persisting) of the pattern we have tracked, orbital motion 
is certainly the simplest explanation.

\newpage

\centerline{\bf REFERENCES}

\begin{enumerate}

\item \label{jacoby92} 
  Jacoby, G.~H. \etal\
  A critical review of selected techniques for measuring extragalactic distances.
  {\sl PASP} {\bf 104}, 599--662 (1992).

\item \label{madore99} 
  Madore, B.~F. \etal\
  The Hubble Space Telescope Key Project on the Extragalactic Distance Scale.
     XV.  A Cepheid distance to the Fornax Cluster and its implications.
  {\sl Astropjys. J.} {\bf 515}, 29--41 (1999).

\item \label{watson94} 
  Watson, W.~D. \& Wallin, B.~K. 
  Evidence from masers for a rapidly rotating disk at the nucleus of NGC4258.
  {\sl Astrophys. J.} {\bf 432}, L35--L38 (1994).

\item \label{miyoshi95} 
  Miyoshi, M. \etal\
  Evidence for a black hole from high rotation velocities in a 
    sub-parsec region of NGC4258.
  {\sl Nature} {\bf 373}, 127--129 (1995).

\item \label{greenhill95a} 
  Greenhill, L.~G., Jiang, R.~D., Moran, J.~M., Reid, M.~J., Lo, K.~Y., \& Claussen, M. J.
  Detection of a subparsec diameter disk in the nucleus of NGC4258.
  {\sl Astrophys. J.} {\bf 440}, 619--627 (1995).

\item \label{herrnstein96}
  Herrnstein, J.~R., Greenhill, L.~J., \& Moran, J.~M.
  The Warp in the subparsec molecular disk in NGC4258 as an explanation for 
     persistent asymmetries in the maser spectrum
  {\sl Astrophys. J.} {\bf 468}, L17--L20 (1996).

\item \label{myphd} 
  Herrnstein, J.~R. 
  PhD Dissertation, Harvard University, 1997.

\item \label{nakai95} 
  Nakai, N., Inoue, M., Miyazawa, K., Miyoshi, M., \& Hall, P. 
  Search for extremely high velocity \ho\ maser emission in Seyfert galaxies.
  {\sl Pub. Astron. Soc. Japan} {\bf 47}, 771--799 (1995).

\item \label{greenhill95b} 
  Greenhill, L.~J., Henkel, C., Becker, R., Wilson, T.~L., \& Wouterloot, J.~G.~A. 
  Centripetal acceleration within the subparsec nuclear maser disk of NGC4258.
  {\sl Astron. \& Astrophys.} {\bf 304}, 21--33 (1995).

\item \label{bragg98} 
  Bragg, A.~E., Greenhill, L.~J., Moran, J.~M., \& Henkel, C.
  Acceleration-derived positions of the high-velocity maser features in NGC4258.
  {\sl BAAS.} {\bf 30}, 1254 (1998).

\item \label{moran95}  
  Moran, J.~M. \etal\
  Probing active galactic nuclei with \ho\ megamasers.
  {\sl Proc. Natl. Acad. Sci. USA} {\bf 92}, 11427--11433 (1995).

\item \label{herrnstein97b} 
  Herrnstein, J.~R. \etal\
  Discovery of a subparsec jet 4000 Schwarzscild radii from the central engine of 
    NGC4258.
  {\sl Astrophys. J.} {\bf 475}, L17--L21 (1997).

\item \label{herrnstein98} 
  Herrnstein, J.~R. \etal\
  VLBA continuum observations of NGC4258: Constraints on an advection-dominated
    accretion flow.
  {\sl Astrophys. J.} {\bf 497}, L69--L73 (1998).

\item \label{cecil95} 
  Cecil, G., Wilson, A.~S., \& DePree, C. 
  Hot shocked gas along the helical jets of NGC4258.
  {\sl Astrophys. J.} {\bf 440}, 181--190 (1995).

\item \label{zensus95} 
  Zensus, J.~A., Diamond, P.~J., \& Napier, P.~J.
  {\sl Very Long Baseline Interferometry and the VLBA} 
  (Astr. Soc. Pacific, San Francisco, 1995).

\item \label{aaronson82}
  Aaronson, M. \etal\ 
  A catalog of infrared magnitudes and HI velocity widths for nearby galaxies. 
  {\sl Astrophys. J. Suppl.} {\bf 50}, 241--262 (1982).
 
\item \label{richter84} 
  Richter, O.-G. \& Huchtmeier, W.~K. 
  Is there a unique relation between absolute (blue) luminosity and total
    21 cm linewidth of disk galaxies?
  {\sl Astron. \& Astrophys.} {\bf 132}, 253--264 (1984).
 
\item \label{rowan85} 
  Rowan-Robinson, M.
  {\sl The Cosmological Distance Ladder} 
  (W.H Freeman and Co., 1985). 
 
\item \label{eracleous95} 
  Eracleous, M., Livio, M., Halpern, J.~P., \& Storchi-Bergmann, T.
  Elliptical Accretion Disks in Active Galactic Nuclei
  {\sl Astrophys. J.} {\bf 438}, 610--622 (1995).

\end{enumerate}

\acknowledgements{The National Radio Astronomy Observatory is a facility of the National
Science Foundation operated under cooperative agreement by Associated 
Universities, Inc.}

\newpage

\centerline{ {\bf CAPTIONS} }

Figure~1. -- The upper panel shows the best-fitting warped disk model superposed on actual
maser positions as measured by the VLBA of the NRAO, with top as North.   The 
filled square marks the center of 
the disk, as determined from a global disk-fitting analysis\cit{myphd}.  The filled 
triangles show the positions of the {\it high-velocity} masers, so called because they occur 
at frequencies corresponding to Doppler shifts of $\sim\pm1000$\,\kms\ with respect to the 
galaxy systemic velocity of $\sim470$\,\kms. This is apparent in the VLBA total power 
spectrum displayed in the lower panel.  The inset shows line-of-sight (LOS) velocity versus 
impact parameter for the best-fitting Keplerian disk, with the maser data superposed.  
The high-velocity masers trace a Keplerian curve to better than 1\%.
Monitoring of these features indicates that they drift by less than 
$\sim1$\,\kmsyr\,\,\cittt{nakai95}{greenhill95b}{bragg98} and requires that they lie 
within 5--10$^{\circ}$ of the midline, the intersection of the disk with the plane of 
the sky. The LOS velocities of the {\it systemic} masers are centered 
about the systemic velocity of the galaxy.  The positions (filled circles of upper panel) 
and LOS velocities of these masers imply they subtend about $8^{\circ}$ of disk azimuth 
centered about the LOS to the central mass, and the observed 8--10\,\kmsyr\ acceleration 
of these features\citt{nakai95}{greenhill95b} unambiguously places them along the near 
edge of the disk.  The approximately linear relationship between systemic maser impact 
parameter and LOS velocity demonstrates that the disk is very thin\cit{moran95} (aspect 
ratio $\la0.2\%$) and that 
these masers are confined to a narrow annulus in the disk.  The magnitude of the velocity 
gradient (\omegas) implies a mean systemic radius, \avgrs, of 
3.9 mas which, together with the positions of the high-velocity masers, constrains the disk 
inclination, \is, to be $\sim82\pm1^{\circ}$ ($90^{\circ}$ for edge-on). Finally, VLBA 
continuum images\citt{herrnstein97b}{herrnstein98} are included as 
contours in the upper panel.  The 22-GHz radio emission traces a sub-parsec-scale jet 
elongated along the rotation axis of the disk and well-aligned with a luminous, kpc-scale 
jet\cit{cecil95}. 

Figure~2. -- Line-of-sight (LOS) velocities (a) and right ascensions (b) at the peaks of 
the systemic maser spectrum for each of the five VLBA epochs. Only those features 
deemed reliably trackable by the pattern-matching analysis are included.  All epochs 
included the Very Large Array, phased to act as a single large aperture.  In 
addition, the final epoch utilized the Effelsberg 100-meter telescope.  In each epoch the 
VLBA correlator provided cross-power spectra with channel spacings of 0.22\,\kms.  The NGC4258 
masers are characterized by linewidths of $\sim1-3$\,\kms.  All spectra were phase-referenced, 
via self-calibration, to a strong systemic maser to stabilize the interferometer against 
atmospheric pathlength fluctuations, and synthesis images were constructed for each spectral 
channel using conventional restoration techniques\cit{zensus95}.  Error bars have
been foregone in order to avoid clutter.  The uncertainties in individual LOS velocity estimates 
are dominated by line blending in the spectrum. There is an average scatter of about 
0.4\,\kms\ about the best-fitting acceleration tracks.  The masers are spatially 
unresolved and relative positions have been measured to a precision of 
$\sim0.5\Theta_{B}/\mbox{SNR}$, where 
SNR is the signal to noise ratio and $\Theta_{B}$ represents the $0.6\times0.9$\,mas, 
approximately North-South synthesized beam. Relative positional accuracies typically
ranged from 0.5 to 10 \muas.   All positions are 
relative to a fixed point along the systemic position-velocity gradient.  Our ability to 
precisely align this structure amongst all epochs suggests that it does indeed remain 
fixed in time.  The best-fitting acceleration and proper motion tracks (solid lines)
indicate average drifts of $9.3$\,\kmsyr\ and $31.5$\,\muasyr\ in velocity and 
position, respectively.  The scatter in the individual proper motions and accelerations 
about these average values is consistent with a 0.2 mas scatter in the radii of the systemic 
masers about \avgrs.  

Figure~3. -- Systemic maser bulk proper motion (\avgpm; a) and acceleration (\avgax; b) 
probability density functions (PDFs) as derived using the Bayesian pattern-matching 
analysis described in the text.  The curves were generated using all the maser features 
in each epoch.  The uncertainties in \avgpm\ and \avgax\ as derived from these PDFs 
include measurement uncertainties in the maser positions and velocities, but they are 
dominated by ambiguities in tracking specific maser features.

\end{document}